\documentclass[journal]{IEEEtran}
\usepackage{url,amsmath,graphicx}
\usepackage{fancyhdr}
\usepackage{cite}
\usepackage{graphicx}
\usepackage{psfrag}
\usepackage{subfigure}
\usepackage{url}
\usepackage{stfloats}
\usepackage{amsmath}
\usepackage{array}
\usepackage{fancyhdr}
\usepackage{epsfig}
\usepackage{amssymb}
\usepackage{color}
\usepackage{csquotes}
\MakeOuterQuote{"}
\newcommand\Exp{\mathbb{E}}
\usepackage{selinput}
\SelectInputMappings{%
	adieresis={ä},
	eacute={é},
	Lcaron={Ľ},
}
\providecommand{\keywords}[1]{{{\textit\textbf Index Terms---}#1}}
\usepackage[font=scriptsize]{caption}

\begin{document}
\title{ Shadowed  FSO/mmWave
Systems with Interference}

\author{ Im\`ene~Trigui, Member, \textit{IEEE}, Panagiotis D. Diamantoulakis, Senior Member, \textit{IEEE},  Sofi\`ene~Affes, Senior Member \textit{IEEE}, and George K. Karagiannidis, Fellow, \textit{IEEE}. }


\maketitle
\begin{abstract}
We investigate the performance of mixed free space
optical (FSO)/millimeter-wave (mmWave) relay networks with interference at the destination.
The FSO/mmWave channels are assumed to
follow M\'alaga-$\mathcal{M}$/ Generalized-$\cal K$ fading
models with pointing errors in the FSO link.
The {\rm H}-transform theory, wherein integral transforms
involve Fox's {\rm H}-functions as kernels, is embodied to unifying the performance analysis framework  that encompasses closed-form expressions  for the outage probability, the average bit error rate (BER) and the average capacity.    By virtue of some {\rm H}-transform asymptotic expansions, the  high  signal-to-interference-plus-noise ratio (SINR) analysis reduces to easy-to-compute expressions for the outage probability and BER, which reveals inside information for the system design.  We
finally investigate  the optimal power allocation strategy, which minimizes  the outage probability.
	\let\thefootnote\relax\footnotetext{Work supported by the Discovery Grants and the CREATE PERSWADE (www.create-perswade.ca) programs of NSERC and a Discovery Accelerator Supplement (DAS) Award from NSERC.
}
\end{abstract}
{\keywords Millimeter wave, dual-hop relaying, free-space optics (FSO),  cochannel interference (CCI), M\'alaga-$\mathcal{M}$ fading, power allocation, shadowing.}
\section{Introduction}

Millimeter wave (mmWave) small-cell concept is envisioned to enable extremely high data rates and ubiquitous coverage through the resources
reuse over smaller areas and the huge amount of available
spectrum. One significant
concern in the deployment of such networks is backhauling in order to handle the unprecedented data traffic surge between all
small cells across the network. Recently, due to its cost-efficient and  high data rate capabilities and immunity to interference, the current  perspectives advocate the use of free-space optics  (FSO) technology as a promising solution for constructing low-cost backhaul for small-cells.  In this perspective, relay-assisted  FSO-based backhaul framework and mmWave-based access links, where relays are
applied as  optical to radio frequency (RF) "converter" to assist the communications of
 small cells, is considered as a powerful candidate to provide  high-data rate reliable communications in high-density heterogeneous networks \cite{xe},\!\! \cite{dehos}.
Nevertheless, several hurdles must be overcome to enable mixed FSO/mmWave communications and make them work properly.  One of the major challenges facing the application of FSO communication is  its vulnerability to atmospheric turbulence and strong path-loss \cite{mal}.  On the  RF side, on the other hand,  the  mmWave signals can be blocked due to shadowing thereby inferring coverage holes that prevent mmWave
communication from delivering uniform capacity for all users
in the network \cite{rice}. Moreover,  in ultra-dense cellular networks,  the mmWave RF interference issue may arise when the signals emitted from  a large number of unintended transmitters are captured by the beam at an intended receiver via line-of-sight (LoS) and/or reflection paths, thereby critically exacerbating the link quality deterioration \cite{int}.

\subsection{State-of-The Art and Motivation}
In  recent years,  understanding the fundamental performance  limits of mixed FSO/RF systems has attracted a lot of  research interest (see \cite{Zedini}-\!\cite{Balti2} and references therein). Also, the effective utilization of resources (e.g., power) in both combined systems becomes of paramount importance. In \cite{Zedini} and \cite{Anees},  the authors investigated the performance of an amplify-and-forward (AF) mixed RF/FSO relay network over Nakagami-$m$ and Gamma-Gamma fading channels. Exact
closed-form and analytical expressions were, respectively,  derived
in \cite{Zedini} and \cite{Anees} for the outage probability,
average bit error rate (BER), and channel capacity. Considering
the outdated channel-state-information (CSI) effect on the RF link
and misalignment error on the FSO link, the authors in
\cite{Djordjevic} evaluated the performance of an AF mixed RF/FSO
relay network over Rayleigh and  Gamma-Gamma fading models. The
same system model was studied in \cite{Zhang}, but with
$\kappa$-$\mu$ and $\eta$-$\mu$ fading models for the RF link and a
Gamma-Gamma fading model for the FSO link. Whereas it was studied in \cite{Kong}
assuming Rayleigh fading  for the RF link and a
M\'alaga-$\mathcal{M}$ distribution model for the FSO link. In \cite{Nasab}, the authors investigated the performance of an AF
mixed RF/FSO relay network while including the direct link
between the source and destination. They assumed Nakagami-$m$
fading model for the RF links and a generalized  Gamma-Gamma fading
model for the FSO link when deriving closed-form expressions for the
outage and bit error probabilities. Work on AF mixed RF/FSO
relay networks continued in \cite{Trinh1} where the authors
considered a millimeter-wave (mmWave) Rician distributed RF channel
and a M\'{a}laga-$\mathcal{M}$ distributed FSO channel. The
same system model was also considered in \cite{Hajji}, while
assuming Weibull and Gamma-Gamma fading models for the mmWave RF
and FSO links, respectively.   In \cite{Trigui}, the
authors studied the performance of a mixed FSO/RF relay network
assuming M\'alaga-$\mathcal{M}$/shadowed $\kappa$-$\mu$ fading
models. They derived exact and asymptotic (i.e., at high signal-to-noise
ratio (SNR) values) closed-form expressions for the system outage
probability and channel capacity. Several studies on the effect of interference on the performance of
AF mixed FSO/RF relay networks  are
presented in \cite{Trigui2}, \cite{Balti2}, and \cite{cherif}. The mixed RF/FSO relay network was investigated in \cite{Lei}
from a security point of view  and in a cognitive radio scenario in \cite{Arezumand}.  The performance
of an AF mixed RF/FSO relay network with
multiple antennas at the source and multiple apertures at the
destination was investigated in \cite{Yang}. Most recently, Balti et al. \cite{Balti1} proposed
a mixed RF/FSO system with general model of hardware
impairments considering optical channels with Gamma-Gamma fading.


Although the results from \cite{Zedini}-\!\cite{Balti2} are insightful, these works have been successfully tractable only for small-scale fading channels on the RF links
or  M\'alaga-$\mathcal{M}$ FSO links. To the  best of our knowledge,
the performance analysis of mixed FSO/mmWave systems
under M\'alaga-$\mathcal{M}$ distribution and composite fading conditions
where fading and shadowing phenomena occur simultaneously has not been investigated in the open literature. In fact,  in mmWave networks,  both the desired and interfering signals are adversely
effected by shadowing  from objects over the signal path due
to high directivity or due to human body  movements \cite{h1}. Shadowing along with the high attenuation  are the main drawbacks at
mmWave frequencies that hinder successful transmission.  As such, a careful characterization of
the mixed FSO/mmWave system over composite fading conditions is crucial to identify the negatives
of higher attenuation and shadowing. However, since composite fading distributions are steadily challenging,  a friendlier
analytical approach that typically allows the derivation of tractable expressions
for key performance measures  and indicators of interest is in fact desirable, yet still missing.
While the work in \cite{Trigui} provides innovative characterization of
mixed FSO/RF relay systems in fading channels where only dominant LOS components  are affected by Nakagami-$m$ distributed  shadowing,
co-channel interference has not been considered. In fact,  the incorporation of RF mmWave interference  has been, so far,  steadily overlooked (e.g., see \cite{Trinh1}, \! \!\!\cite{Yang}) in the mixed FSO/RF context.

In this paper, we tackle the above issues by providing holistic
analytical tools facilitating the evaluation of the  mixed FSO/mmWave
relay network performance  by considering general
cases, i.e.,  shadowed  small-scale fading both on the desired and
interference links, which are more challenging to analyze than only including
distance-dependent path loss or rayleigh fading \cite{Balti1}-\!\cite{heath}.
\subsection{Technical Contribution}

In this paper, we investigate the performance of a dual-hop mixed
FSO/mmWave relay network.  To model mmWave composite multi-path shadowing fading, we consider
generalized-$\cal K$ distribution (\!\!\cite{miridakis2},\!\cite{imenet},\! \cite{bithas}) with parameters $m$ and $\kappa$  where different $m$ values  represent
LOS and NLOS cases \cite{heath} and $\kappa$ indicates  the mmWave sensitivity to
blockages.  To mitigate the effects of multi-path fading, the relay-to-destination mmWave-based hop uses a multiple-input-single-output (MISO) setup with $N$ transmit antennas.  We further  assume that the FSO link undergoes M\'alaga-$\mathcal{M}$  distribution.
 Furthermore, it is assumed that the destination is affected by independent identically distributed co-channel interference in the mmWave band. The contribution of this paper can be summarized as follows:

\begin{itemize}
  \item Using the theory of Fox's H-functions and Mellin–Barnes
integrals, we propose a novel mathematical framework to derive closed-form expressions for important statistics of the SINR under the assumption of fixed and variable-gain
relaying,  while not making any
assumptions in our derivations, in terms of the bivariate Fox's H
function.
  \item New analytical results  for the outage probability, the average error probability, and average capacity are derived.  Our  analysis procedure  and performance metrics formulations are given in unified and
tractable mathematical fashion thereby serving  as a useful tool
to validate and compare the special cases of M\'alaga-$\mathcal{M}$ and generalized-$\cal K$ distributions.

  \item An asymptotic  outage and error rate performance analysis is presented,  which enables the characterization of the key
performance indicators, such as the diversity gain and
coding gain, size of transmit  array, effect of pointing error and shadowing on the achieved performance under the presence of interference.

  \item Capitalizing on the achieved asymptotic results, the optimum relay power allocation that minimizes the system outage
probability  is derived.

\end{itemize}

\subsection{Organization}
The remainder of this paper is organized as follows. We
describe the system and channel models in Section II. In Section III, we present
 the unifying H-transform analysis of the end-to-end SINR statistics for both fixed-gain
and channel-state-information (CSI)-assisted mixed FSO/mmWave networks.  Then, in section IV, we derive exact closed-form expressions for the outage probability,
the average error probability and the average capacity followed by their
asymptotic expressions obtained at high SINR. In section V,  the optimum design strategy
for FSO/mmWave networks is studied.  Section VI presents some numerical and simulation results to
illustrate the mathematical formalism presented in the previous
sections. Finally, some concluding remarks are drawn out in
Section VII.
\section{System Model}
We consider the downlink of a relay-assisted network featuring a mixed FSO/mmWave communication link as shown in Fig.~1. The source $S$ is assumed to include
a single photo-aperture, while the relay node $R$ is assumed to have a
single photo detector from one side and $N$ antennas from the
other side. The relay is able to activate either heterodyne or intensity modulation/direct (IM/DD) detection.
Using amplify-and-forward (AF) relaying,  all the $N$ transmit antennas at the relay are used for MRT (maximum ratio transmission) to  communicate with the destination $D$  over the mmWave band.
   In the first hop, the FSO signal undergoes a  M\'alaga-$\mathcal{M}$
turbulent-induced fading channel, while in the second hop, the mmWave signals undergoes a generalized-$\cal K$
fading channel. We further assume that the destination  is affected by  $L$ interferers.
The interferers affecting $D$ have independent identically distributed generalized-$\cal K$ fading.

\begin{figure}[t!]
	\centering
	\includegraphics[scale=0.3]{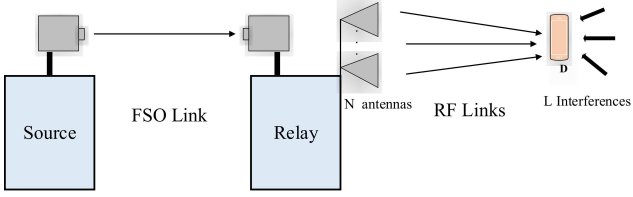}
	\caption{A dual-hop interference-limited mixed FSO/mmWave RF relay system.}
	\label{fig:fif}
\end{figure}

\subsection{Optical Channel Model}
The FSO ($S$-$R$) channel  follows a M\'alaga-$\mathcal{M}$  distribution for which the cumulative density function (CDF) of the  instantaneous SNR $\gamma_1$  in the presence of pointing errors is given  by \cite[Eq. (5)]{ansari}
\begin{eqnarray}
	\label{eq:9}
	F_{\gamma_1}(x)&=&\frac{\xi^2A r}{\Gamma(\alpha)}\sum_{k=1}^{\beta}\frac{b_k}{\Gamma(k)}\nonumber\\
	&&{\rm \mathcal{H}}_{2,4}^{3,1}\Biggl[\frac{B^rx}{\mu_r}\Bigg\vert{ (1,r),(\xi^2+1,r) \atop(\xi^2,r),(\alpha,r),(k,r),(0,r)}\Biggr],
\end{eqnarray}
where  $\xi$ is the ratio between the equivalent beam radius and the pointing error displacement standard deviation (i.e., jitter) at the relay (for negligible pointing errors $\xi \rightarrow+\infty$),
$ A={ \alpha^{\frac{\alpha}{2}}\left[  {g \beta }/({g \beta +\Omega})\right] ^{\beta +\frac{\alpha}{2}}}{g^{-1-\frac{\alpha}{2}}}$ and $ {b_k}\!\!=\!\binom{\beta\!-\!1}{k\!-\!1}\!{\left(g \beta\!+\Omega \right)^{1-\frac{k}{2}}}\left[ {(g \beta +\!\Omega)}/{\alpha\beta}\right] ^{\frac{\alpha+k}{2}}\left( {\Omega}/{g}\!\right)^{k-1}\left({\alpha}/{\beta}\!\right)^{\frac{k}{2}}$, where $\alpha$, $\beta$, $g$  and $\Omega$ are the fading parameters related to the atmospheric turbulence conditions \cite{ansari}, \cite{navas2}. It may be useful to mention that $g=2b_{0}(1-\rho)$ where $2 b_{0}$ is the average power of the LOS term and $\rho$ represents the amount of scattering power coupled to the LOS component ($0  \leqslant \rho_i  \leqslant 1 $)\footnote{It is worth highlighting that the $\mathcal{M}$ distribution unifies most of the proposed statistical models characterizing  the optical irradiance in  homogeneous and isotropic turbulence \cite{ansari}. Hence both $\mathcal{G}$-$\mathcal{G}$ and ${\cal K}$ models are special cases of the M\'alaga-$\mathcal{M}$ distribution, as they mathematically derive from (\ref{eq:9})  by setting ($g=0$, $\Omega=1$) and ($g\neq0$, $\Omega=0$ or $\beta=1$), respectively \cite{ansari}.}. Moreover in (\ref{eq:9}), ${\rm \mathcal{H}}_{p, q}^{m, n} [\cdot]$ and $\Gamma(\cdot)$ stand for the Fox's-{\rm H} function \cite[Eq. (1.2)]{mathai} and the  Gamma function \cite[Eq. (8.310.1)]{grad}, respectively, and $B={\alpha\beta h (g+\Omega)}/(g\beta+\Omega)$ with  $h=\xi^2/(\xi^2+1)$. Furthermore, $r$ is the parameter that describes the detection technique at the relay (i.e., $r = 1$
is associated with heterodyne detection and $r = 2$ is associated
with IM/DD) and $\mu_r$ refers to the electrical SNR of the FSO
hop \cite{ansari}. In particular, for $r = 1$,
\begin{equation}
\mu_{1}=\mu_{\text{heterodyne}}=\Exp[\gamma_1]=\bar{\gamma}_1,
\end{equation}
and for $r=2$,  it becomes \cite[Eq.(8)]{ansari}
\begin{equation}
\mu_{2}=\mu_{\text{IM/DD}}= \frac{\mu_{1}\alpha\xi^2(\xi^2+1)^{-2}(\xi^2+2)(g+\Omega)}{(\alpha+1)[2g(g+2\Omega)+\Omega^2(1+\frac{1}{\beta})]}.
\end{equation}

\subsection{MmWave Channel Model}
MmWave signals are extremely sensitive to objects, including foliage and human body. Shadowing effect in the
mmWave communication comes then to prominence. In this paper, we consider a complete channel model with shadowing, path loss and small-scale fading. As such, we express the $X$-$D$, $X\in\{R, I\}$  channel in the following form
 \begin{equation}
 {\bf h}_{XD}=\sqrt{P_X \psi(d_{XD})}\widetilde{{\bf h}}_{XD},
 \label{hx}
 \end{equation}
  where  $\widetilde{{\bf h}}_{XD}=\{\widetilde{ h}_{XD,1},\ldots,\widetilde{ h}_{XD,\delta_X}\}$ captures the effects of small-scale fading with $\delta_X=\{N,L\}$ for $X\in \{R,I\}$, and  $\psi(d_{XD})$ captures the effect of large-scale fading on ($X$-$D$) links, and $P_X$ is the power of the signal transmitted from $X$ to $D$.
$\widetilde{{\bf h}}_{XD}$ is assumed to follows  Nakagami-$m_X$  where  $m_X$, $X\in \{R,I\}$ indicates the
degree of fading severity. In mmWave LOS links, the number of scatterers is
relatively small. Thus, the LOS link fading is less severe, which
is modeled by relatively large $m_X$\footnote{Though the modeling of LOS  mmWave-based links is well known for line-of-sight wireless links with Rice fading \cite{rice}, the latter  can be well approximated  by the
Nakagami-$m$ model with parameter $m_{X}=\frac{\left(K_X+1\right)^{2}}{2 K_X+1}$, where $K_X$, $X\in \{R,I\}$, is the Rician factor.}. Conversely, the NLOS
parameter $m_X$ is smaller \cite{heath}. Therefore, several works (ex.,  \cite{Trinh1}, \cite{heath}, \cite{chelli}) have
suggested Nakagami-$m$ fading, a general yet tractable
model for mmWave bands. It should be noted that accurate
cluster-based channel models such as the Saleh-Valenzuela
model \cite{ven} are mathematically intractable. Thus, we omit
such models in this work.
Hereafter, we use the shorthand notation for the RV $Z \sim {\cal G}(\alpha, \beta)$
to denote that $Z$ follows a Gamma distribution
with parameters $\alpha$ and $\beta$. From (\ref{hx}), we have the total small-scale received signal/interference at the destination
$Y_{XD}=\sum_{i=1}^{\delta_X}\widetilde{h}^{2}_{XD,i}$ is the sum of $\delta_X$ independent identically distributed (i.i.d.) Gamma RVs $\widetilde{h}^{2}_{XD,i}\sim  {\cal G}(m_{X}, \frac{1}{m_X})$.
It can easily be shown that $Y_{XD}$ is also Gamma distributed with parameters $\delta_X m_{X}$ and $1/m_X$, i.e.,
$Y_{XD}  \sim {\cal G}(\delta_X m_{X}, \frac{1}{m_X})$. From (\ref{hx})  the  root-mean-square power of the received
signal is subject to variations induced by shadowing and path loss. Then, under the assumption of generalized-$\cal K$ model \cite{bithas} and to  capture the shadowing effects, we use  a Gamma distribution with  parameter $\kappa_{xd}$ i.e., $\psi(d_{xd})\sim {\cal G}(\kappa_{X}, \bar{\gamma}_{X}/\kappa_{X})$, where $\kappa_X\geqslant 0$ denotes the
shadowing severity and $\bar{\gamma}_{X}=P_{X}{\cal E}\{\psi(d_{XD})\}$ where $\cal E(\cdot)$ is the expectation operator.
It is demonstrated that  the corresponding PDF  of the instantaneous SNR (respectively  INR), $\gamma_{XD}=\sum_{i=1}^{\delta_X}h^{2}_{XD,i}$,  $X\in (R, I)$, is  given by \cite[Eq. (5)]{imenet},\cite[Eq. (9.34.3)]{grad} as
\begin{eqnarray}
	\label{eqC2:4bis}
	f_{\gamma_{XD}}(x)&=&\frac{\frac{m_X \kappa_X}{\bar{\gamma}_{X}}}{\Gamma(\delta_Xm_X)\Gamma(\kappa_X)}\nonumber\\
	&&{\rm G}_{0,2}^{2,0}\Biggl[\!\frac{\kappa_X\!m_X}{\bar{\gamma}_{X}} x\Bigg\vert \  {-\atop \delta_Xm_X\!-\!1,\kappa_X-1}\Biggr],
\end{eqnarray}
 where ${\rm G}_{p, q}^{m, n} [\cdot]$  stands for the Meijer's-G  \cite[Eq. (9.301)]{grad}  function.
 The term $\bar{\gamma}_{X}=P_{X}{\cal E}\{\psi(d_{xd})\}$  represents the average received
power for the link between  $X\in \{R, I\}$ and the destination.
The CDF of the signal-to-interference ratio (SIR)  $\gamma_2=\gamma_{RD}/\gamma_{ID}$ under $\cal GK$ fading can be derived from a recent result in \cite[Lemma 1]{miridakis2} as
\begin{eqnarray}
	\label{eqC2:5}
	F_{\gamma_2}(x)&=&1-\frac{1}{\Gamma(Nm)\Gamma(\kappa)\Gamma(Lm_I)\Gamma(\kappa_I)}\nonumber\\
	&&{\rm G}_{3,3}^{3,2}\Biggl[\frac{\kappa m x}{\kappa_Im_I\bar{\gamma}}\Bigg\vert  {1-\kappa_I,1-Lm_I,1 \atop 0,\kappa,Nm}\Biggr],
\end{eqnarray}
where  $\bar{\gamma}=\bar{\gamma}_{RD}/\bar{\gamma}_{ID}$ is the average SIR of the RF link where, for consistency, we have dropped the subscript $R$ from the parameters $m_R$ and $\kappa_R$.
 Path loss models for mmWave signals have been proposed
in \cite{mmv} and \cite{mmv1} for $28$ GHz and $38$ GHz, respectively. Using
these models, we can express the path loss experienced by the
signal in the ($X$-$D$) link as
\begin{equation}
20\log_{10}\left( \frac{4 \pi d_0}{\lambda_{W}}\right)+10\eta \log_{10}\left(\frac{d_{XD}}{d_0}\right),
\label{path}
\end{equation}
 where $d_{XD}$ refers to the distance between the relay/interference and the destination, $d_0$ is a free-space reference distance set to 5 meters in  \cite{mmv}, \cite{mmv1}, $\lambda_{W}$ stands for the wavelength ($7.78$ mm in $38$ GHz and
$10.71$ mm in $28$ GHz)\footnote{The $28$ GHz is one of the standardized bands for the $5$G cellular operation
\cite{mmv1}.}, and $\eta$ stands for the path-loss exponent.
MmWave channel measurements in \cite{mmv} and \cite{mmv1} have shown that
the value of the path-loss exponent $\eta$ is equal to $2.2$ in $38$ GHz
and $2.55$ in $28$ GHz. Using the path loss model for mmWaves
in (\ref{path}), we can express the average received power
over the $X$-$D$ hop as
\begin{equation}
\bar{\gamma}_{X}=P_{X} \left(\frac{ \lambda_W}{4 \pi d_0}\right)^{2}\left(\frac{d_0}{d_{XD}}\right)^{\eta}.
\end{equation}
Recently, there have been convincing measurements revealing that  mmWave channels are often dominated by both the LOS and  first-order reflection paths \cite{int}. In such environments, it is possible that any LoS and/or reflection components from
surrounding interferers can critically deteriorate the link quality, thus increasingly biasing the system towards interference-limited regime as  BS and user densities increase \cite{heath}, \cite{chelli}.
While many-element adaptive arrays can boost the received signal power and hence reduce
the impact of interference  \cite{heath}, characterizing the accumulated
 interference  from a large number of unintended transmitters still plays an important role in evaluating and
predicting the dense mmWave networks performance.


 In this work, under the assumption of interference-limited mmwave links, we express the end-to-end SINR of mixed FSO/mmwave system for
fixed-gain relaying as \cite[Eq. (6)]{Zedini}
\begin{equation}
	\label{eqC2:8}
	\gamma=\frac{\gamma_1\gamma_2}{\gamma_2 +{\cal C}},
\end{equation}
where $\gamma_2\triangleq \gamma_{RD}/\gamma_{ID}$ is defined as the RF interference-to-noise ratio (INR)  and  $\cal C$ stands for the fixed gain at the relay. On the other hand, the end-to-end SINR when CSI-assisted relaying scheme is considered is
 expressed  as \cite[Eq. (7)]{Zedini}
 \begin{equation}
 \label{eq:10}
 \gamma=\frac{\gamma_1 \gamma_2}{\gamma_1+\gamma_2+1}.
 \end{equation}
In what follows, we derive analytical expressions for key performance metrics of mixed FSO/mmWave dual-hop systems for both kinds of relay amplification schemes.

\section{Performance Analysis of Fixed-Gain Relaying}
\label{sec:3}
Under the assumption of interference-limited regime and  considering fixed-gain relaying, exact and asymptotic expressions for the
outage probability and the error rate probability are proposed.

 \textit{Theorem 1 (Exact Outage Probability) :}
	The outage probability is defined as
the probability that the end-to-end SINR falls below predetermined threshold $\gamma_{th}$ and is obtained as
	\begin{equation}
		\label{eqC2:outfix1}
		P_{\text{out}}=F_\gamma(\gamma_{th}),
	\end{equation}
where
\begin{eqnarray}
	\label{eqC2:CDFfix}
	F_\gamma(x)&=&\frac{\xi^2A \kappa m {\cal C} }{\Gamma(\alpha)\Gamma(Nm)\Gamma(\kappa)\Gamma(Lm_I)\Gamma(\kappa_I)\kappa_Im_I\bar{\gamma}}\nonumber\\
	&&\!\!\!\!\!\!\!\!\sum_{k=1}^{\beta}\frac{b_k}{\Gamma(k)}{\rm \mathcal{H}}_{1,0:3,2:4,5}^{0,1:0,3:4,3}\left[{ \frac{\mu_r}{B^rx}\atop \frac{\kappa m{\cal C}}{\kappa_Im_I\bar{\gamma}}} \left|\begin{array}{cccc}(0,1, 1) \\-\\(\delta,\Delta)\\(\lambda,\Lambda)\\(\chi, X)\\(\upsilon,\Upsilon) \end{array}\right.\right],
\end{eqnarray}
where  ${\rm H}^{m_1,n_1:m_2,n_2:m_3,n_3}_{p_1,q_1:p_2,q_2:p_3,q_3}[\cdot]$ denotes the Fox-H function of two variables\cite[Eq. (1.1)]{mittal} whose Mathematica implementation may be found in \cite[Table I]{Lei}, whereby $(\delta,\Delta)=(1-\xi^2\!,r),(1-\alpha,r),(1-k,r)$; $(\lambda,\Lambda)=(0,1),(-\xi^2,r)$; $(\chi, X)=(-1,1),(-\kappa_I,1),(-Lm_I,1),(0,1)$; and $(\upsilon,\Upsilon)=(-1,1),(-1,1),(\kappa-1,1),(Nm-1,1),(0,1)$.
\begin{IEEEproof}
See Appendix A.
\end{IEEEproof}
The PDF of the end-to-end SINR $\gamma$ for shadowed FSO/mmWave systems  is obtained  as
\begin{eqnarray}
	\label{eqC2:15}
	f_\gamma(x)&=&-\frac{\xi^2A \kappa m {\cal C}  }{x\Gamma(\alpha)\Gamma(Nm)\Gamma(\kappa)\Gamma(Lm_I)\Gamma(\kappa_I)\kappa_Im_I\bar{\gamma}}\nonumber\\
	&&\!\!\!\!\!\!\!\!\!\sum_{k=1}^{\beta}\frac{b_k}{\Gamma(k)}{\rm \mathcal{H}}_{1,0:3,2:4,5}^{0,1:0,3:4,3}\left[{ \frac{\mu_r}{B^rx}\!\atop \frac{\kappa m{\cal C}}{\kappa_Im_I\bar{\gamma}}} \left|\begin{array}{cccc}(0,1, 1) \\-\\(\delta,\Delta)\\(\lambda',\Lambda')\\(\chi, X)\\(\upsilon,\Upsilon) \end{array}\right.\right],
\end{eqnarray}	
where $(\lambda',\Lambda')=(1,1),(-\xi^2,r)$.
\begin{IEEEproof}
	The result follows from  differentiating  the Mellin-Barnes integral in (\ref{eqC2:CDFfix}) over $x$ using $\frac{dx^{-s}}{dx}=-sx^{-s-1}$ with $\Gamma(s+1)=s\Gamma(s)$ and applying \cite[Eq. (2.57)]{mathai}.
\end{IEEEproof}

In the effort to understand the impact of key parameters on
outage  performance, we look into the asymptotic
regime in the high optical SNR $\widetilde{\mu}_r$ and RF SIR ${\bar \gamma} \rightarrow \infty$, based on which the diversity
and coding gains are obtained.

 \textit{Lemma 1 (Asymptotic Outage probability):}
\label{cor:1}
	At high normalized average SNR  in the FSO link ($\frac{\mu_r}{\gamma_{th}}\rightarrow\infty$), the outage probability of the system under consideration is obtained as
	\begin{eqnarray}
		P_{\text{out}}&\underset{\frac{\mu_r}{\gamma_{th}}\gg1}{\approx}&\frac{A \xi^2 }{r\Gamma(\alpha)\Gamma(Nm)\Gamma(\kappa)\Gamma(Lm_I)\Gamma(\kappa_I) }\sum_{k=1}^{\beta} \frac{b_k}{\Gamma(k)} \nonumber\\
		&&\!\!\!\!\!\!\!\!\!\!\!\!\!\!\!\!\!\!\!\!\!\!\!\!\!\!\!\Biggl( \Lambda \left(\!\frac{\kappa m{\cal C}B^r\gamma_{th}}{\kappa_Im_I\bar{\gamma}\mu_r}\!\right)^{\min\{N m,\kappa, \frac{\xi^2}{r},\frac{\alpha}{r},\frac{k}{r}\}}\!+\!\frac{\Gamma(\alpha\!-\!\xi^2)\Gamma(k\!\!-\!\xi^2)}{ \Gamma(1\!-\!\frac{\xi^2}{r})}\nonumber\\
		&&\!\!\!\!\!\!\!\!\!\!\!\!\!\!\!\!\!\!\!\!\!\!\!\!\!\!\!~\Xi\!\left(\!\gamma_{th},\frac{\xi^2}{r}\!\right)+\!\frac{\Gamma(\xi^2\!\!-\!\!\alpha)\Gamma(k\!\!-\!\alpha)}{ \Gamma(1\!-\!\frac{\alpha}{r})\Gamma(1\!+\!\xi^2\!-\!\alpha)}~\Xi\left(\!\gamma_{th},\frac{\alpha}{r}\!\right)\nonumber\\
		&&\!\!\!\!\!\!\!\!\!\!\!\!\!\!\!\!\!\!\!\!\!\!\!\!\!\!\!+\frac{\Gamma(\xi^2\!\!-\!k)\Gamma(\alpha\!-\!k)}{ \Gamma(1\!-\!\frac{k}{r})\Gamma(1\!+\!\xi^2\!-\!k)}~\Xi\left(\!\gamma_{th},\frac{k}{r}\!\right)\Bigg),
			\label{eqC2:outfix1high}
\end{eqnarray}
	where
 \begin{equation}
\Xi(x,y)\!=\!\left(\!\frac{B^rx}{\mu_r}\!\right)^{y}\!\!{\rm G}_{3,3}^{3,3}\!\!\left[\!{\frac{\kappa m{\cal C}}{\kappa_I m_I\bar{\gamma}}}\!\!\left|  \begin{array}{cccc}\!\!\!1\!-\kappa_I\!,1\!-\!Lm_I\!,1+y\!\\ \kappa,Nm,0\end{array}\!\!\!\!\right.\right],
\end{equation}
   and $\Lambda$ is a constant.

\textit{Proof:}
	The proof of the above result is given in Appendix B  with the use of  the asymptotic expansion of $\mathcal{H}$-function
\cite[Eq. (1.8.7)]{mathai}
\begin{equation}
 {\mathcal H}_{p,q}^{m,n}\left[ x \left|\begin{array}{ccc} (a_i, A_j)_p \\ (b_i,B_j)_q \end{array}\right. \right]\underset{x\rightarrow0}{\approx} \Lambda x^{c},
\end{equation}
where $c=\underset{j=1,\ldots, m}{\min}\left[\frac{\mathfrak{R} (b_j)}{B_j}\right]$, and $\Lambda$ is given in \cite[Eq. (1.8.5)]{mathai}.

  With the aim of obtaining the diversity order and
coding gain of the system,  the CDF in (\ref{eqC2:outfix1high}) can be simplified
at the high SNR values to be
\begin{eqnarray}
		P^{\infty}_{\text{out}} &\underset{{\bar \gamma} \gg1}{\approx}&(\mathcal{G}_c {\bar \gamma} )^{-{\cal G}_d}\nonumber \\
&\!\!\!\!\!\!\!\!\!\approx\!\!\!\!\!\!\!\!\!& \frac{A \xi^2 }{r\Gamma(\alpha)\Gamma(Nm)\Gamma(\kappa)\Gamma(Lm_I)\Gamma(\kappa_I) }\sum_{k=1}^{\beta} \frac{b_k}{\Gamma(k)} \nonumber\\
		&& \Lambda \left(\!\frac{\kappa m{\cal C}B^r\gamma_{th}}{\kappa_Im_I\bar{\gamma}\mu_r}\!\right)^{\min\{N m,\kappa, \frac{\xi^2}{r},\frac{\alpha}{r},\frac{k}{r}\}} ,
					\label{eqC2:outfix1high1}
\end{eqnarray}
 where  $\mathcal{G}_d$ stands for the diversity gain and is defined
as the slope of the asymptotic curve, and  $\mathcal{G}_c$ is the coding
gain representing the SNR advantage of the asymptotic curve
relative to ${\bar \gamma}_k ^{-G_d}$ reference.
 From (\ref{eqC2:outfix1high1}), it can be deduced that the outage probability of the system can be reduced
by increasing the SIR at the FSO and RF links. Moreover, (\ref{eqC2:outfix1high1}) implies that the outage performance is governed by the hop
that has the worst propagation condition for the desired signal, whereas the number of interferers has no impact on the diversity gain. Numerical results in
Section VI show that the approximation in (\ref{eqC2:outfix1high}) and (\ref{eqC2:outfix1high1}) are very tight
at high SIR.  As a special case,   the diversity gain under
Gamma-Gamma turbulence is obtained from (\ref{eqC2:outfix1high1}) as
 \begin{equation}
  {\cal G}_d=\min\left(N m, \kappa, \frac{\xi^2}{r},\frac{\alpha}{r},\frac{\beta}{r}\right),
  \label{div}
  \end{equation}
   while the achievable  coding
gain can be expressed as
\begin{eqnarray}
{\cal G}_c&=& \frac{\kappa_Im_I}{\kappa m{\cal C}B^r\gamma_{th}}\nonumber \\ &&\!\!\!\!\!\!\!\!\!\!\!\!\!\!\!\!\!\!\!\!\!\!\!\!\!\left(\frac{\Lambda  \xi^2 }{r\Gamma(\alpha)\Gamma(\beta)\Gamma(Nm)\Gamma(\kappa)\Gamma(Lm_I)\Gamma(\kappa_I)}\right)^{\!\!\!-\frac{1}{\!\!\min\{N m,\kappa, \frac{\xi^2}{r},\frac{\alpha}{r},\frac{\beta}{r}\}}}.\nonumber \\
\end{eqnarray}

\textit{Theorem 2 (Exact Error Probability):}
The end-to-end error probability is obtained as
\begin{eqnarray}
		\label{eqC2:fix5}
		\mathcal{B}&=&\frac{\xi^2A \varphi\kappa m {\cal C}}{2\Gamma(\alpha) \Gamma(p)\Gamma(Nm)\Gamma(\kappa)\Gamma(Lm_I)\Gamma(\kappa_I)\kappa_Im_I \bar{\gamma}} \nonumber\\
		&&\!\!\!\!\!\!\!\!\!\!\!\!\!\!\!\!\!\!\!\!\sum_{j=1}^{n}\sum_{k=1}^{\beta}\frac{b_k}{\Gamma(k)}{\rm \mathcal{H}}_{1,0:3,3:4,5}^{0,1:1,3:4,3}\!\!\left[{ \frac{\mu_rq_j}{B^r}\atop \frac{\kappa m{\cal C}}{\kappa_Im_I\bar{\gamma}}} \!\left|\!\!\begin{array}{cccc}(0,1, 1) \\-\\(\delta,\Delta)\\(p,1),(\lambda,\Lambda)\\(\chi, X)\\(\upsilon,\Upsilon) \end{array}\right.\!\!\!\right].
	\end{eqnarray}	
\begin{IEEEproof}
The average BER
can be written in terms of the CDF of the end-to-end SIR  as
\begin{equation}\label{eqC2:ber1}
\mathcal{B}=\frac{\varphi}{2\Gamma(p)}\sum_{j=1}^{n}q_j^p\int_{0}^{\infty}e^{-q_jx}x^{p-1}F_{\gamma}(x)\mathrm{d}x,
\end{equation}
where $\Gamma(\cdot,\cdot)$ stands for the incomplete Gamma function \cite[Eq. (8.350.2)]{grad} and the parameters $\varphi$, $n$, $p$ and $q_j$ account for different modulations schemes \cite{imenet}.
Now, substituting the Mellin-Barnes integral form  of (\ref{eqC2:CDFfix}) using \cite[Eq. (2.56)]{mathai} into (\ref{eqC2:ber1}) and resorting to \cite[Eq. (7.811.4)]{grad}, we obtain  (\ref{eqC2:fix5}) after some manipulations.
\end{IEEEproof}	

\textit{Lemma 2 (Asymptotic Error Probability):}
At high normalized average SNR  in the FSO link ($\frac{\mu_r}{\gamma_{th}}\rightarrow\infty$), the asymptotic average BER  is derived  as
	\begin{eqnarray}
		\label{eqC2:berfix1high}
		\mathcal{B}^{\infty}&\underset{\mu_r\gg1}{\approx}&\frac{\xi^2A \varphi\kappa m {\cal C}}{2\Gamma(\alpha) \Gamma(p)\Gamma(Nm)\Gamma(\kappa)\Gamma(Lm_I)\Gamma(\kappa_I)\kappa_Im_I \bar{\gamma}} \nonumber\\
		&&\!\!\!\!\sum_{j=1}^{n}\sum_{k=1}^{\beta}\frac{b_k}{\Gamma(k)}\Biggl[\frac{\Gamma(\alpha-\xi^2)\Gamma(k-\xi^2)}{r \Gamma(1-\frac{\xi^2}{r})}\Xi\left(\frac{1}{q_j}\frac{\xi^2}{r}\right) \nonumber\\
		&&\!\!\!\!+\frac{\Gamma(\xi^2-\alpha)\Gamma(k-\alpha)}{r \Gamma(1-\frac{\alpha}{r})\Gamma(1+\xi^2-\alpha)}\quad\Xi\left(\frac{1}{q_j},\frac{\alpha}{r}\right) \nonumber\\
		&&\!\!\!\!+\frac{\Gamma(\xi^2-k)\Gamma(\alpha-k)}{r \Gamma(1-\frac{k}{r})\Gamma(1+\xi^2-k)}\quad\Xi\left(\frac{1}{q_j},\frac{k}{r}\right) \nonumber\\
		&&\!\!\!\!+\frac{B^r}{\mu_rq_j}{\rm H}_{4,5}^{5,3}\Biggl[{\frac{\kappa m{\cal C}B^r}{\kappa_Im_I\bar{\gamma}\mu_rq_j}}\left|\begin{array}{cccc}(\sigma',\Sigma')\\(\phi,\Phi)\end{array}\right.\!\!\! \Biggr]\!\Biggr],
	\end{eqnarray}
	where $(\sigma',\Sigma')=(-\!\kappa_I,1),(-\!Lm_I,1),(-p,1),(1+\xi^2-r,r)$.
	\begin{IEEEproof}
		The asymptotic error probability follows  along the same lines of Appendix B, while resorting to the Fox's  $\mathcal{H}$ function asymptotic expansion
 in (\ref{eqC2:berfix1high})  yields  a similar result to (\ref{eqC2:outfix1high}).
	\end{IEEEproof}
\textit{Theorem 3 (Average Capacity):}
	The average capacity of the considered mixed FSO/RF mmWave relaying system under heterodyne detection technique  can be computed as  $2 \ln(2)\mathcal{C_E}={\mathcal E}\left\{\ln(1+\gamma)\right\}$, thereby yielding
	\begin{eqnarray}
		\label{eqC2:fix21}
\mathcal{C_E}&=&\frac{\xi^2A \kappa m {\cal C} }{2 \ln(2)\Gamma(\alpha)\Gamma(Nm)\Gamma(\kappa)\Gamma(Lm_I)\Gamma(\kappa_I)\kappa_Im_I\bar{\gamma}} \nonumber\\
		&&\!\!\!\!\!\!\!\!\!\!\!\!\!\!\sum_{k=1}^{\beta} \frac{b_k}{\Gamma(k)}{\rm \mathcal{H}}_{1,0:4,3:4,5}^{0,1:1,4:4,3}\left[{ \frac{\mu_r}{B^rx}\atop \frac{\kappa m{\cal C}}{\kappa_Im_I\bar{\gamma}}}\left|\begin{array}{cccc}(0,1, 1) \\-\\(\delta,\Delta),(1,1)\\(0,1)(\lambda',\Lambda')\\(\chi, X)\\(\upsilon,\Upsilon) \end{array}\right.\right].
	\end{eqnarray}	  	 	 	 	
\begin{IEEEproof}
Averaging $\ln(1+\gamma)={\rm G}_{2,2}^{1,2}\biggl[ \gamma \bigg\vert\!\!\ {{1,1} \atop {1,0}}\biggr]$ over the end-to-end SINR PDF obtained from differentiating  (\ref{eqC2:CDFfix}) while resorting to \cite[Eq. (1.1)]{mittal} and \cite[Eq. (7.811.4)]{grad} yields the result after  some  manipulations.
\end{IEEEproof}

\textit{Remark 1:} The M\'alaga-$\mathcal{M}$ reduces to Gamma-Gamma fading when  ($g=0$, $\Omega=1$), whence  all terms in (\ref{eq:9}) vanish except for
the term when  $k=\beta$.
Hence, when $g=0$, $\Omega=1$, $\kappa, \kappa_I \rightarrow \infty$, (\ref{eqC2:fix21}) reduces, when $r=1$,  to the ergodic capacity of mixed Gamma-Gamma FSO/interference-limited Nakagami-$m$ RF transmission with heterodyne detection as given by
	\begin{eqnarray}
		\label{eqC2:fixGGN}
		\mathcal{C_E}&=&\frac{\xi^2}{2 \ln(2)\Gamma(Nm)\Gamma(Lm_I)\Gamma(\alpha)\Gamma(\beta)}{\rm G}_{1,0:4,3:4,3}^{1,0:1,4:3,2}\nonumber\\
		\!\!\!\!\!\!&&\!\!\!\!\!\!\!\!\!\!\!\!\!\!\!\!\Biggl[\!\!\frac{\mu_1}{\alpha\beta h}; \frac{ m{\cal C}}{m_I\bar{\gamma}}\Bigg\vert\ {\!1\atop-\!}\!\ \!\Bigg\vert\!\ {1\!-\!\xi^2,1\!-\!\alpha,1\!-\!\beta,1\atop 1,0,-\xi^2}\!\Bigg\vert \ {1\!-\!Lm_I,1,0\atop Nm,0,1}\!\!\Biggr],\nonumber\\
	\end{eqnarray}
where ${\rm G}_{a,[c,e],b,[d,f]}^{p,q,k,r,l}[\cdot,\cdot]$  is   the generalized Meijer's G-function and is used to represent the product of three Meijer's-G
functions in  closed-form  \cite{verma}.

\textit{Remark 2:} In IM/DD-based optical systems, the signal is constrained to be nonnegative and real-valued. Thus, the input signal distribution to approach Shannon channel capacity does not necessarily follow Gaussian distribution in optical wireless channels. Assuming solely an average optical power constraint and ignoring pre-detection noise at the optical receiver, which is due to random intensity fluctuations of the optical source and shot noise caused by the ambient light, \cite[Eq. (35)]{Zedini}, \cite[Eq. (35)]{r1} can be used where $\mathcal{C_E}\geq{\mathcal E}\left\{\ln(1+ \frac{e}{2\pi}\gamma)\right\}$, which follows in the same line of (\ref{eqC2:fix21}).  This assumption is quite reasonable in our case, since the impact of  thermal noise and RF interference at the receiver, is much higher than pre-detection noise at the optical receiver.

\section{Performance Analysis of CSI-assisted Relaying}
Due to the  intractability  of the SINR in (\ref{eq:10}),  we present in the following subsection new upper bound
expressions for the outage and error rate probabilities.
The SINR in (\ref{eq:10})  can be upper bounded using the standard approximation $\gamma\cong\min\{\gamma_{1},\gamma_{2}\}$. The cumulative distribution function (CDF) of $\gamma$ can be written as
\begin{equation}
F_{\gamma}(\gamma)= 1-\prod_{X\in\{1, 2\}}F_{\gamma_X}^{(c)}(\gamma).
\label{out1}
\end{equation}
The expressions of $F^{(c)}_{\gamma_{X}} (\gamma_{th})$,  $X\in\{1, 2\}$ are already obtained in \cite[Eq.(8)]{Trigui} and (\ref{eqC2:5}). Then, recognizing that  the product of two Fox's $\mathcal{H}$ functions is also a Fox's $\mathcal{H}$ function in (\ref{out1}) yields
\begin{eqnarray}
\label{eq:poutsci}
F_{\gamma}(\gamma)&=&1-\frac{\xi^2Ar}{\Gamma(\alpha)\Gamma(Nm)\Gamma(\kappa)\Gamma(Lm_I)\Gamma(\kappa_I)}\nonumber\\
&&\!\!\!\!\!\!\!\!\!\!\!\!\!\!\!\!\!\!\!\sum_{k=1}^{\beta}\frac{b_k}{\Gamma(k)}{\rm \mathcal{H}}_{0,0:2,4:3,3}^{0,0:4,0:3,2} \left[{ \frac{B^r\gamma}{\mu_r}\!\atop \frac{\kappa m\gamma}{\kappa_Im_I\bar{\gamma}}} \!\left|\begin{array}{cccc}(0,1, 1\!) \\-\\(\delta_1,\Delta_1)\\(\lambda_1,\Lambda_1)\\(\chi_1, X_1)\\(\upsilon_1,\Upsilon_1) \end{array}\right.\right],
\end{eqnarray}
where $(\delta_1,\Delta_1)=(\xi^2+1,r),(1,r)$, $(\lambda_1,\Lambda_1)=(0,r),(\xi^2,r),(\alpha,r),(k,r)$, $(\chi_1, X_1)=(1-\kappa_I,1),(1-Lm_I,1),(1,1)$, and $(\upsilon_1,\Upsilon_1)=(0,1),(\kappa,1),(Nm,1)$.\\
Up to now, the outage probability can be obtained by replacing
$\gamma$ by $\gamma_{th}$ in (\ref{eq:poutsci}).

 With the aim of obtaining the diversity order and
coding gain of the system,  the outage probability  in (\ref{eq:poutsci}) can be simplified
at the high SIR values to be
	\begin{eqnarray}
	\label{eq:pouthighcsi}
{P^{\infty}_{\text{out}}}& \approx &\frac{\xi^2A }{\Gamma(\alpha)\Gamma(Nm)\Gamma(\kappa)\Gamma(Lm_I)\Gamma(\kappa_I)}\nonumber \\ && \sum_{k=1}^{\beta}\frac{b_k}{\Gamma(k)}\!\sum_{j=1}^{5}\frac{\zeta_j}{\Psi_j} \left( {\frac{\gamma_{th}}{\bar{\gamma}}}\right)^{\Psi_j},
	\end{eqnarray}
where $\Psi=\{N m, \kappa, \frac{\xi^2}{r},\frac{\alpha}{r},\frac{k}{r}\}$, $\zeta_1=-\left(\!\!\frac{m\kappa}{m_I \kappa_I}\!\!\right)^{\!\!N m}\Gamma(\kappa N m)\Gamma(\kappa_I\!+\!N m)\Gamma(L m_I\!+\!N m)$,~  $\zeta_2=-\left(\frac{m\kappa}{m_I\kappa_I}\!\!\right)^{\kappa}\Gamma(Nm-\kappa)\Gamma(\kappa_I+\kappa)\Gamma(Lm_I+\kappa)$, $ \zeta_3=\Gamma(\alpha-\xi^2)\Gamma(k-\xi^2)\frac{B^{\xi^2}}{r}$, $\zeta_4=(\xi^{2}-\alpha)^{-1}\Gamma(k-\alpha)\frac{B^{\alpha}}{r}$,  and $\zeta_5=(\xi^{2}-k)^{-1}\Gamma(\alpha-k)\frac{B^{k}}{r}$.

\begin{IEEEproof}
The result in (\ref{eq:pouthighcsi}) follows easily after applying the asymptotic expansion of the Fox-$\rm H$ function given in \cite[Theorem 1.11]{kilbas} to (\ref{eq:poutsci}).
\end{IEEEproof}

In the context of ${P^{\infty}_{\text{out}}} \approx (\mathcal{G}_c {\bar \gamma} )^{-{\cal G}_d}$, it can be inferred from (\ref{eq:pouthighcsi}) that
 \begin{eqnarray}
	\label{eq:pouthighcsi1}
{P^{\infty}_{\text{out}}}&\approx& \frac{\xi^2  }{\Gamma(\alpha)\Gamma(Nm)\Gamma(\kappa)\Gamma(Lm_I)\Gamma(\kappa_I)\Gamma(\beta)}\nonumber \\ &&\sum_{j=1}^{5}\frac{\zeta_j}{\Psi_j} \left( {\frac{\gamma_{th}}{\bar{\gamma}}}\right)^{\min \{N m, \kappa, \frac{\xi^2}{r},\frac{\alpha}{r},\frac{\beta}{r}\}}.
	\end{eqnarray}
It is to be noted that at high SIR regime the lower-bound
of the outage probability provided by (\ref{eq:poutsci}) has the same slope as the exact
outage  in (\ref{eqC2:CDFfix}).


\textit{Lemma 3 (Error Probability):}
	The
error rate probability under CSI-assisted relaying  is obtained as
	\begin{eqnarray}
	\label{eq:17}
	{\mathcal B}&=&\frac{\varphi n}{2}-\frac{ \xi^2A r\varphi
	}{2\Gamma(p)\Gamma(\alpha)\Gamma(Nm)\Gamma(\kappa)\Gamma(Lm_I)\Gamma(\kappa_I)}\nonumber\\
	\!\!\!\!\!\!\!\!\!\!\!\!\!\!\!\!\!\!\!\!\!\!\!\!\!&&\!\!\!\!\!\!\!\!\!\!\!\!\!\!\!\!\!\!\!\!\!\!\!\!\sum_{j=1}^{n}\!\sum_{k=1}^{\beta}\frac{b_k}{\Gamma(k)} {\rm \mathcal{H}}_{1,0:2,4:3,3}^{0,1:4,0:3,2}\left[{\frac{B^r}{\mu_rq_j}\atop \frac{\kappa m}{\kappa_Im_I\bar{\gamma}q_j}}\left| \begin{array}{cccc}(1-p,1, 1) \\-\\(\delta_1,\Delta_1)\\(\lambda_1,\Lambda_1)\\(\chi_1, X_1)\\(\upsilon_1,\Upsilon_1) \end{array}\right.\right].
	\end{eqnarray}  	
\begin{IEEEproof}
	Substituting (\ref{out1}) into (\ref{eqC2:ber1}) and resorting to  \cite[Eq. (1.59)]{mathai} and \cite[Eq. (2.2)]{mittal} yield the result  after some manipulations.
\end{IEEEproof}

\textit{Lemma 4 (Exact Average Capacity): }
	The average capacity of the considered mixed FSO/interference-limited mmWave system  under CSI-assisted relaying and heterodyne detection is expressed by
	\begin{eqnarray}
	\label{eq:18}
\mathcal{C_E}&=&\frac{\xi^2A  r  \mu_r}{2 \ln(2)\Gamma(\alpha)\Gamma(Nm)\Gamma(\kappa)\Gamma(Lm_I)\Gamma(\kappa_I)B^{r}} \nonumber\\
	&&\!\!\!\!\!\!\sum_{k=1}^{\beta} \frac{b_k}{\Gamma(k)} {\rm \mathcal{H}}_{1,0:4,3:3,4}^{0,1:1,4:3,3} \left[{\frac{\mu_r}{B^r}\!\atop \frac{\kappa_Im_I\bar{\gamma}}{\kappa m}} \left| \begin{array}{cccc}(0,1, 1) \\-\\(\delta_2,\Delta_2)\\(\lambda_2,\Lambda_2)\\(\chi_2, X_2)\\(\upsilon_2,\Upsilon_2) \end{array}\right.\right],
	\end{eqnarray}
	where $(\delta_2,\Delta_2)=(1\!-r\!,r\!),(1\!-\xi^2\!-r,\!r)\!, \!(1\!-\!\alpha\!-r,\!r)\!,\!(1\!-\!k\!-r,\!r)\!$, $(\lambda_2,\Lambda_2)=(1,1),(\!1-\kappa,1),(\!1-\!Nm,1)$, $(\chi_2, X_2)=(1,1),(\!1-\kappa,1),(\!1-\!Nm,1)$, and $(\upsilon_2,\Upsilon_2)=(1,1),(\kappa_I,1),(Lm_I,1),(0,1)$.
\begin{IEEEproof}
	See Appendix C.
\end{IEEEproof}
	It should be mentioned that when $r=1$ and  $\kappa, \kappa_I \rightarrow \infty$,  (\ref{eq:18})  reduces to  the ergodic capacity over mixed FSO/inteference-limited mmWave systems in M\'alaga/Nakagami-$m$ fading channels  with heterodyne detection as given by
	\begin{eqnarray}
	\label{eq:15bis}
	\mathcal{C_E}&=&\frac{\xi^2A   \mu_1}{2 \ln(2)B\Gamma(\alpha)\Gamma(Nm)\Gamma(Lm_I)\alpha\beta h}\sum_{k=1}^{\beta}\! \frac{b_k}{\Gamma(k)} \nonumber\\
	&&\!\!\!\!\!\!\!\!\!\!\!\!\!\!\!\!\!\!\!\!\!\!\!\!\!\!\!\!{\rm G}_{1,0:4,3:2,3}^{1,0:1,4:2,2}\Biggl[\!\!\frac{\mu_1}{\alpha\beta h}; \frac{m_I\bar{\gamma}}{m}\Bigg\vert \ {1\atop-}\ \Bigg\vert\!\ \!\!{0,\!-\xi^2,\!-\alpha,\!-k\atop 0,-\xi^2\!-\!1,\!-1}\!\Bigg\vert \ \!\! {\!1,1\!-\!Nm\atop 1,Lm_I,0}\!\!\Biggr].
	\end{eqnarray}
\section{FSO/Mmwave Systems Optimum Design}
This section addresses the optimum resource allocation
strategy at the source  and the relay devices such that the $P_{\text{out}}$ is minimized subject to a sum power constraint.
The total power $P_T$ is equal to the sum of the electrical power
 $P_{\mathcal F}$ assigned to the optical source device and the power
$P_{\cal R}$ assigned to the relay, i.e., $P_{T}=P_{\cal F}+P_{\cal R}$.  To this end, recall that $\bar{\gamma}=\frac{P_{\cal R} \left(\frac{ \lambda_W}{4 \pi d_0}\right)^{2}\left(\frac{d_0}{d_{XD}}\right)^{\eta}}{\bar{\gamma}_I}$.  Moreover, according  to the Beer-Lambert  law  \cite{lam}  the optical beam
power has an exponential decay with propagation distance with $\mu_{r}= P_{\cal F} e^{- \delta d_{\cal F}}$
where $\delta$ is the overall attenuation coefficient. Yet, depending on the accessible emission limits for IM/DD transceivers, $P_{\cal F}$  will be restricted so it does not exceed a power value
of $\mathcal{S}$ Watts.  The optimization problem is then  formulated as follows:
\begin{equation}
\begin{matrix}
\displaystyle \min_{P_{\cal F},P_{\cal R} } &~ P_{\text out} = G  (\mathcal{A}_F P_{\cal F}^{-a}+ \mathcal{A}_R P_{\cal R}^{-a}) \\ \\
\textrm{s.t.}& P_{\cal F}+P_{\cal R} \leq  P_{tot}  \\ \\
& -P_{\cal R}\leq 0, ~ P_{\cal F}\leq \mathcal{S} &
\end{matrix}
\end{equation}
 where $a=\min\{N m,\kappa, \frac{\xi^2}{r},\frac{\alpha}{r},\frac{\beta}{r}\}$, $G=\frac{\xi^2 \mathcal{\gamma}_{th}^{a} }{\Gamma(\alpha)\Gamma(Nm)\Gamma(\kappa)\Gamma(Lm_I)\Gamma(\kappa_I)\Gamma(\beta)}$, $\mathcal{A}_R=\frac{\bar{\gamma}_I^{a}}{\left(\frac{ \lambda_W}{4 \pi d_0}\right)^{2 a}\left(\frac{d_0}{d_{XD}}\right)^{a \eta}}(\zeta_1+\zeta_2)$, and  $\mathcal{A}_F= e^{a \delta d_{\cal F}}(\zeta_3+\zeta_4+\zeta_5)$.
The optimum design of the considered system follows from differentiating the Lagrange cost function \cite{lag}: $\eta_L=P_{\text out}+\delta_L (P_{\cal F}+P_{\cal R}-P_{tot})$ where $\delta_L$ is the Lagrange parameter with respect of the desired parameter $P_{ X}$, $X\in\{\cal F, \cal R\}$ and $\delta_L$, and solving the obtained equations equaled to zero.
Hence, the optimum power allocation subject to
sum power constraint is derived as
\begin{equation}
P^{*}_{ X}=\frac{A_{ X}^{b}}{A_{\cal F}^{b}+A_{\cal R}^{b}}P_{tot}, \quad X\in\{\cal F, \cal R\},
\label{outopt}
\end{equation}
where $b=\frac{1}{a+1}$. From (\ref{outopt}), it can be deduced that the optimal
power $P^{*}_{ R}$ increases if (i) the interference level $\bar{\gamma}_I$ affecting the mmWave signal
 rises, or (ii) the power attenuation due to the distance
travelled by the signal is larger for the mmWave hop compared to
the FSO hop.

\section{Numerical results}
\label{sec:4}
In this section, numerical examples are shown to substantiate
the accuracy of the new unified mathematical framework and to confirm
its potential for analyzing mixed FSO/mmWave communications.  Next, we validate our analysis by comparing the analytical
results with Monte-Carlo simulations\footnote{The results for the Monte-Carlo simulations are obtained by using
100 million samples.}.
The following analysis is conducted in  different shadowing scenarios ranging form infrequent light shadowing ($\kappa=75.5$)
to frequent heavy shadowing ($\kappa=1.09$). The corresponding standard
deviations $\sigma$ of the Lognormal shadowing are equal, respectively, to $0.5$ and  $3.5$ dB   by a moment matching technique
 given by   $\kappa=\frac{1}{e^{\sigma^{2}}-1}$ \cite{imenet}. Unless specified otherwise, Table 1 lists all
the simulation parameters adopted in what follows, which are employed in various FSO and mmWave communication systems \cite{int}, \cite{Trinh1}, \cite{Balti1}, \cite{chelli}.
\begin{table}\caption{\\System and Channel Parameters}
\begin{equation}
\begin{array}{ccc}\\ \hline\hline\\
   \textbf{Parameter} & \textbf{Value} \\ \hline\\
 \text{MmWave bandwidth} & 28 ~\text{GHz}\\ \hline\\
  \text{Reference distance} ~(d_0)& 5~\text{m}\\ \hline\\
\text{Path loss exponent}~ (\eta) & 2.5\\  \hline\\
\text{Relay fixed Gain} ~({\cal G}) & 1.7 \\  \hline\\
\text{Relay antenna number} ~(N) & 2 \\  \hline\\
\text{Attenuation parameter} ~(\delta) & 0.5 \\ \hline\\
\text{Moderate turbulence}~ (\alpha,\beta) &  (5.4, 3.8) \\ \hline\\
\text{Strong turbulence} ~ (\alpha,\beta)&  (2.4, 1.7) \\ \hline\\
 \end{array}
\nonumber
\end{equation}
\end{table}

\begin{figure}[!t]
	\centering\begin{minipage}{.46\textwidth}
	\includegraphics[width=\textwidth]{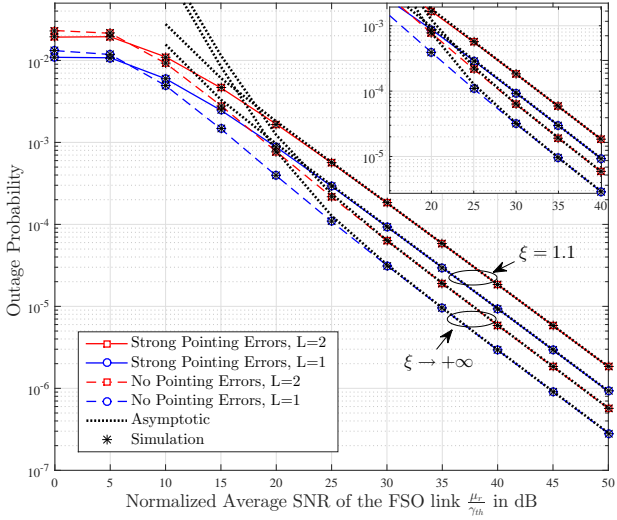}
	\caption{The outage probability  of fixed-gain AF FSO/mmWave relaying system with  IM/DD technique ($r=2$) for different number of interferers in moderate turbulence and frequent heavy shadowing ($\kappa=1.09$) when  $N=2$, $m=m_I=2.5$, $\kappa_I=3.5$, and $\bar{\gamma}=20$ dB.}
	\label{fig:outfixedinterference}\end{minipage}
\end{figure}

 Fig.~\ref{fig:outfixedinterference} depicts the outage probability of fixed-gain   mixed FSO/interference-limited mmWave systems with $L=\{1,2\}$  in frequent heavy shadowed environment ($\kappa=1.09$) versus  the  FSO link normalized average SNR. As expected, increasing $L$ deteriorates the system performance, by increasing the outage probability whereas the diversity gain remains unchanged.  Actually, it can be deduced
from (\ref{div}) that the slope of the outage probability at high SNR depends only
on the fading and turbulence parameters and is not affected
by the number of interferers $L$. Yet, under severe shadowing, a strong pointing error impairment with $\frac{\xi^{2}}{r}>\kappa$ has no effect on the outage diversity gain. Therefore, it is natural that we obtain the same slope for the outage curves even if the value of $\xi$ varies. From Fig.~\ref{fig:outfixedinterference},  it can be observed that the asymptotic expansion in  (\ref{eqC2:outfix1high}) matches very well its exact counterpart at high SNRs.

\begin{figure}[!t]
	\centering\begin{minipage}{.46\textwidth}
	\includegraphics[width=\textwidth]{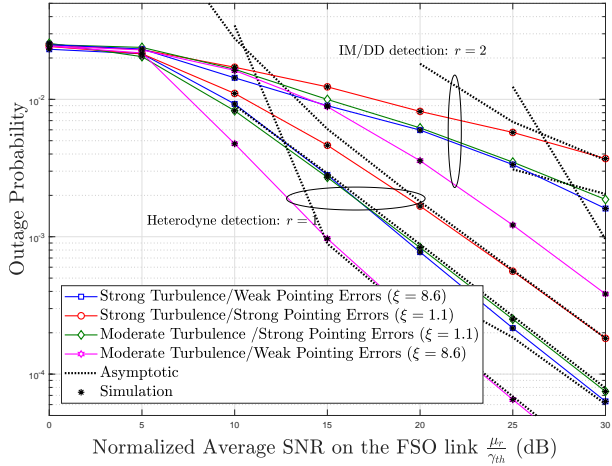}
	\caption{The outage probability of fixed-gain AF FSO/interference-limited frequent heavy shadowed mmWave system under different turbulence and pointing errors severities  with $N=L=2$,  $m=m_I=2.5$ and $\kappa_I=3.5$.}
	\label{fig:Pout_Fixed_Turbulence_PE}\end{minipage}
\end{figure}

Fig.~\ref{fig:Pout_Fixed_Turbulence_PE} illustrates the outage probability of mixed FSO/interference-limited frequent heavy shadowed mmWave versus the  FSO link normalized average SNR in strong  and moderate turbulence conditions, respectively.
As expected, the outage probability deteriorates by decreasing the pointing error displacement standard deviation, i.e., for smaller $\xi$, or decreasing the turbulence fading parameter, i.e., smaller $\alpha$ and $\beta$.
 It is observed that the simulation
results are in excellent agreement with the derived  exact and asymptotic expressions
in  (\ref{eqC2:CDFfix})  and  (\ref{eqC2:outfix1high1}) thereby indicating their accuracy.  The behaviour
of the outage probability can be categorized into two types. Under IM/DD detection, we have ${\cal G}_d=\frac{\xi^{2}}{2}< \kappa$ under strong pointing errors and  ${\cal G}_d=\frac{\beta}{2}< \kappa$ under weak pointing errors and strong turbulence. Otherwise (i.e., $r=1$ and/or weak pointing errors and moderate turbulence), we have ${\cal G}_d=\kappa=1.09$.
Therefore, in this case,  as expected   we obtain the same slope for the outage curves even if the value of $\xi$, $\alpha$, and $\beta$ vary  with increasing SNR since
the effect of mmWave link becomes dominant.

\begin{figure}[!t]
	\centering\begin{minipage}{.46\textwidth}
	\includegraphics[width=\textwidth]{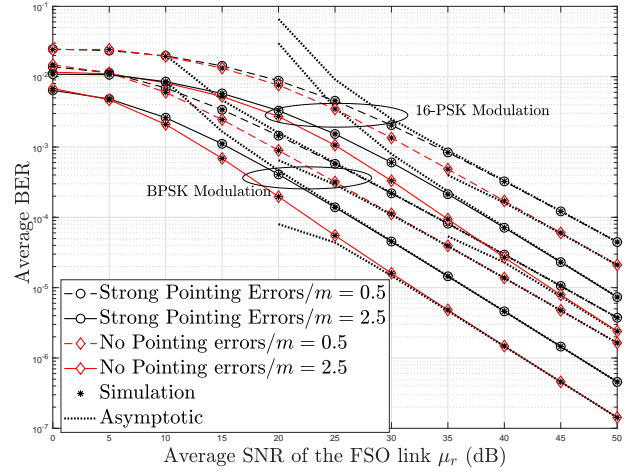}
	\caption{The average BER  of an interference-limited  fixed-gain mixed FSO/mmWave system for heterodyne technique ($r=1$) against the average SNR on the FSO link in strong turbulence conditions and frequent heavy shadowing ($\kappa=1.09$) under varying $m$  with $N=L=2$, $m_I=2.5$, and  $\kappa_I=3.5$.}
	\label{fig:berfixed}\end{minipage}
\end{figure}

Fig.~\ref{fig:berfixed}  depicts the average BER of dual-hop FSO/interference-limited mmWave systems
using fixed-gain relaying for BSPK and 16-PSK modulation schemes  over moderate and strong
 pointing error conditions.  In our numerical
examples, we use large  and small values of the fading parameter $m$ to represent  the LOS ($m=0.5$) and  NLOS ($m=2.5$) conditions, respectively. We observe that  severe fading in the
mmWave link ($m = 0.5$) diminishes the system performance and this
degradation is greater when the FSO link undergoes negligible  pointing errors. The asymptotic results for the average BER at high
SNR on the FSO link derived in Eq. (22) are also included in Fig.~\ref{fig:berfixed} showing an
excellent  tightness at high SNR regime.

\begin{figure}[!t]
	\centering
\begin{minipage}{.46\textwidth}
	\includegraphics[width=\textwidth]{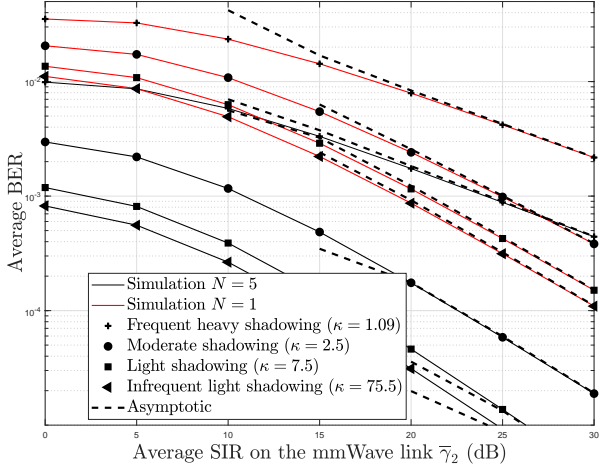}
	\caption{The exact and asymptotic average BER  of an interference-limited  fixed-gain mixed RF/FSO system with  heterodyne technique ($r=1$) under different shadowing scenarios when $L=2$,  $m=1.5$, $m_I=1.5$, and $\kappa_I=3.5$. }
	\label{fig:berfixedshad}
\end{minipage}
\end{figure}

Fig.~\ref{fig:berfixedshad}  demonstrates the average BPSK BER performance of fixed-gain   mixed FSO/interference-limited mmWave systems under several shadowing conditions on the mmWave link, while assuming strong turbulence regime on the FSO link  with fixed effect
of the pointing error ($\xi = 7.1$). A general observation is that the shadowing degrades the system's overall performance. Moreover,  it can be observed,  except for heavy shadowing with $\kappa=1.09$, that all the BEP curves  have the same slopes, which is natural since the BEP ant high SNR/SIR  depends only on the minimum value  ${\mathcal G}_d=\min\left(N m, \kappa, \frac{\xi^2}{r},\frac{\alpha}{r},\frac{k}{r}\right)$. For the two curves when $\kappa=1.09$, they have the same slope revealing equal diversity order ${\mathcal G}_d=\kappa$.
According to Fig.~\ref{fig:berfixedshad},  spatial diversity resulting from employing a higher number of antennas  $N$ at the relay enhances the overall system
performance. Fig.~\ref{fig:berfixedshad} also shows that the asymptotic expansion in  (\ref{eqC2:berfix1high}) agrees very well with the simulation results, hence corroborating its accuracy.

%
\begin{figure}[!t]	\centering
	\begin{minipage}{.46\textwidth}
	\includegraphics[width=\textwidth]{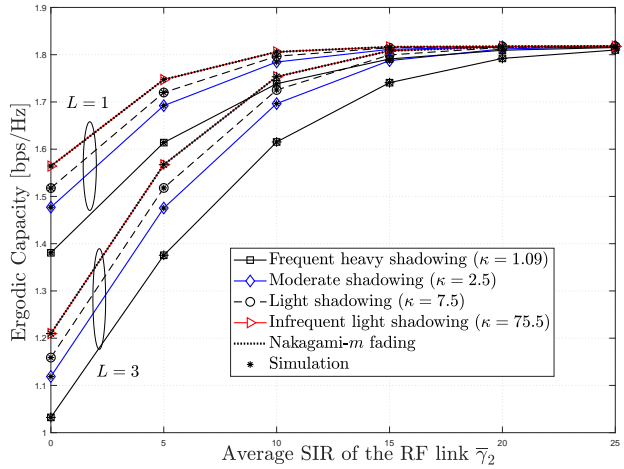}
	\caption{The ergodic capacity of an interference-limited  CSI-assisted AF mixed FSO/mmWave system for different number of interferers $L$  in heavy, moderate, and light shadowing  conditions
 with $N=2$, $m=m_I=2.5$, and $\kappa_I=1.09$.}
	\label{fig:capfixed}\end{minipage}
\end{figure}

	Fig.~\ref{fig:capfixed} investigates the effect of shadowing severity  on the ergodic capacity of mixed FSO/mmWave  CSI-assisted relaying suffering interference. A general observation is that the shadowing degrades the system's overall performance. Furthermore, it can be inferred from Fig.~\ref{fig:capfixed}
that as the SIR of the mmWave link increases, a negligible effect of shadowing and interference on
the capacity is observed and the performance remains almost the
same since the weaker link acts as the dominant link, which is
the FSO link in this case. This can be explained by (25). It may be also useful to mention that the ergodic capacity curves of mixed FSO/mmWave under infrequent light shadowing  ($\kappa\rightarrow\infty$) and mixed M\'alaga-$\mathcal{M}$/Nakagami-$m$ systems coincides thereby unambiguously corroborating the much wider scope claimed by our novel  analysis framework and the rigor of its mathematical derivations.

%

\begin{figure}[!t]	\centering
	\begin{minipage}{.46\textwidth}
	\includegraphics[width=\textwidth]{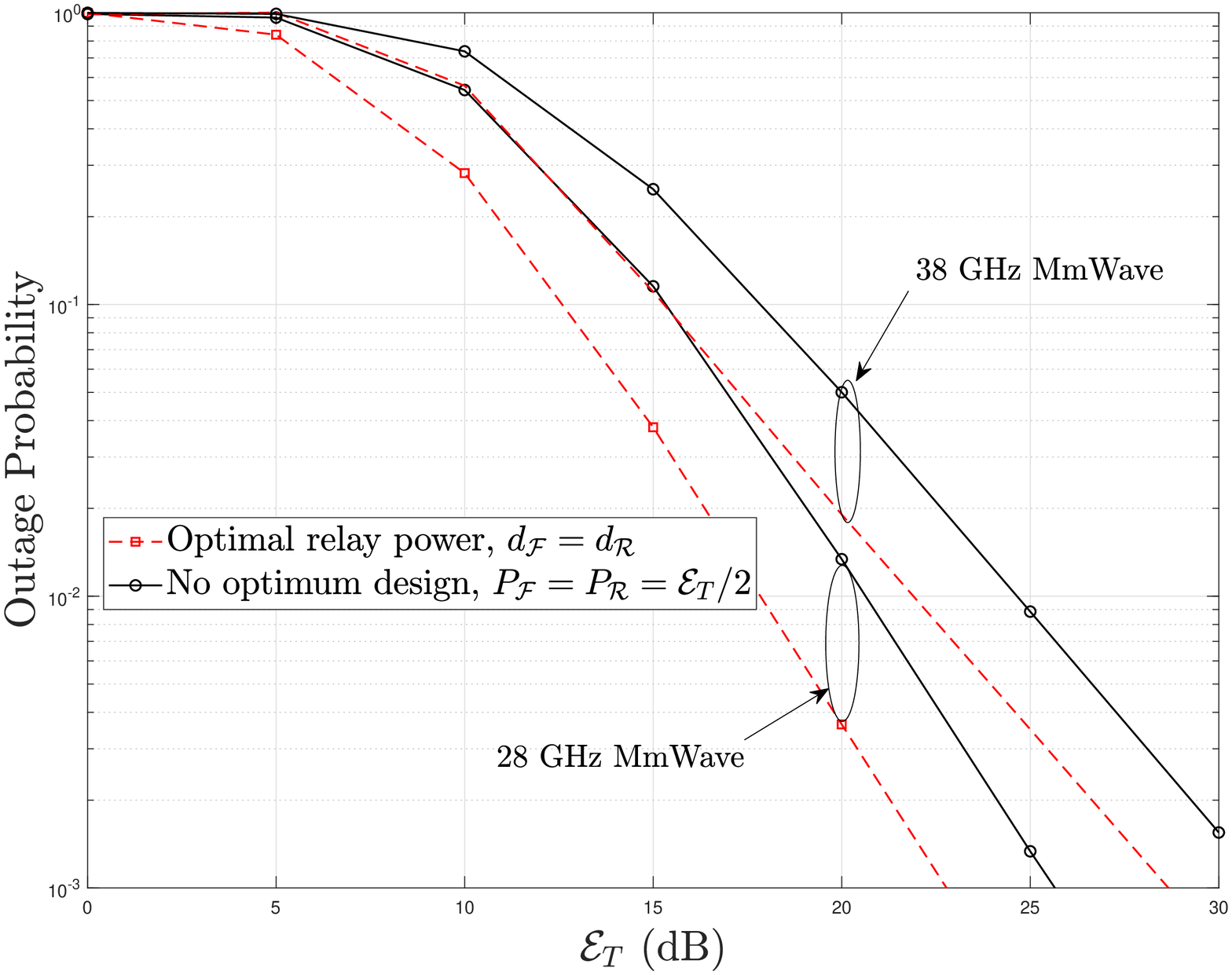}
	\caption{The outage probability with optimal power allocation for different
mmWave bands under moderate turbulence and strong pointing errors on the FSO link for heterodyne technique ($r=1$) and frequent heavy shadowing  on the mmWave links with $m=m_I=2.5$.
when $L=3$.}
	\label{fig:popt}\end{minipage}
\end{figure}
 			
 	Fig.~\ref{fig:popt} shows the impact of power allocation on
the outage probability of  mixed FSO/mmWave relay system against $P_{\text tot}={\cal E}_T$ dB when $\gamma_{th} = 5$ dB and $\gamma_{I} = 2$ dB.
 Moreover, we investigate the impact of the proposed  power allocation
formula in (\ref{outopt})  on the outage performance  and
compare it then to the baseline scheme with no power allocation, i.e., $P_{\cal F}=P_{\cal R}={\cal E}_T/2$,   over different mmWave bandwidths. It can
be observed that the outage decreases with optimal power
allocation compared than with equal power allocation. The achieved gain is of $3.5$ dB   at  a target outage of $10^{-2}$.
  It can be seen from Fig.~\ref{fig:popt} that
the outage decays as the mmWave bandwidth decays.

\section{Conclusion}
\label{sec:5}
We have studied  the performance of relay-assisted
mixed FSO/mmWave  systems with RF interference and shadowing.
The {\rm H}-transform theory is involved into a unified performance analysis framework featuring  closed-form expressions for the
outage probability, the BER and the average capacity assuming M\'alaga-$\mathcal{M}$/generalized-${\cal K}$
channel models for the FSO/shadowed mmWave links while taking into account pointing errors. The diversity order and coding
gain are derived for all studied scenarios.  Furthermore,  we derived an analytical expression for the
optimal power allocation at each hop. Main results showed that under weak atmospheric turbulence
conditions, the system performance is dominated by
the RF channels and a diversity order of $N m$ is achieved by the
system in light shadowing. Otherwise  diversity order is affected by the minimum value of the
turbulence fading, light shadowing,  and pointing error parameters.
\section{Appendix A\\Proof of Theorem 1}
\label{app:A}
The CDF of the end-to-end SINR $\gamma$ with fixed-gain relaying scheme can be derived as
\begin{equation}
	\label{eqC2:apA1}
	F_\gamma(x)=\displaystyle \int_{0}^{\infty} F_{\gamma_1}\left( x\left( \frac{{\cal C}}{y}+1\right) \right) f_{\gamma_2}(y)\mathrm{d}y,
\end{equation}		
where  $F_{\gamma_1}$ and $f_{\gamma_2}$ are the FSO link's CDF and the RF link's PDF, respectively.
 Differentiation of (\ref{eqC2:5}) over $x$ yields $f_{\gamma_2}$  as
\begin{eqnarray}
	\label{eqC2:PDFRF}
	f_{\gamma_2}(x)&=&\frac{-\kappa m}{\Gamma(Nm)\Gamma(\kappa)\Gamma(Lm_I)\Gamma(\kappa_I)\kappa_I m_I \bar{\gamma}}\nonumber\\
	&&\!\!\!\!\!\!{\rm G}_{4,4}^{3,3}\Biggl[\frac{\kappa mx}{\kappa_Im_I\bar{\gamma}}\Bigg\vert  {-1,-\kappa_I,-Lm_I,0 \atop -1,\kappa-1,Nm-1,0}\!\Biggr].
\end{eqnarray}
Substituting (\ref{eq:9}) and (\ref{eqC2:PDFRF}) into (\ref{eqC2:apA1}) while resorting to the integral representation of the Fox-H \cite[Eq. (1.2)]{mathai} and Meijer-G \cite[Eq. (9.301)]{grad} functions yields
\begin{eqnarray}
	\label{eqC2:14}
	F_\gamma(x)&=&\frac{-\xi^2Ar\kappa m }{\Gamma(\alpha)\Gamma(Nm)\Gamma(\kappa)\Gamma(Lm_I)\Gamma(\kappa_I)\kappa_I m_I \bar{\gamma}} \nonumber\\
	&&\!\!\!\!\!\!\!\!\!\!\!\!\!\!\!\!\!\!\!\!\!\!\!\!\!\!\! \sum_{k=1}^{\beta}\frac{b_k}{\Gamma(k)}\frac{1}{4\pi^2i^2}\int_{\mathcal{C}_1}^{}\int_{\mathcal{C}_2}^{}\frac{\Gamma(\xi^2+rs)\Gamma(k+rs)\Gamma(\alpha+rs)}{\Gamma(\xi^2+1+rs)\Gamma(1-rs)}\nonumber\\
	&&\!\!\!\!\!\!\!\!\!\!\!\!\!\!\!\!\!\!\!\!\!\!\!\!\!\!\!\times \frac{\Gamma(-rs)\Gamma(-1-t)}{\Gamma(1+t)} \frac{\Gamma(\kappa-1-t)\Gamma(N m-1-t)}{\Gamma(-t)}\nonumber\\
	&&\!\!\!\!\!\!\!\!\!\!\!\!\!\!\!\!\!\!\!\!\!\!\!\!\!\!\!\times\Gamma(2+t)\Gamma(1+\kappa_I+t)\Gamma(1+Lm_I+t) \left(\!\frac{\kappa m}{\kappa_Im_I\bar{\gamma}}\!\right)^{t}\nonumber\\
	&&\!\!\!\!\!\!\!\!\!\!\!\!\!\!\!\!\!\!\!\!\!\!\!\!\!\!\! \left(\!\frac{B^r x}{\mu_r}\!\right)^{\!\!-s}\int_{0}^{\infty}\left( 1+\frac{\cal C}{y}\right) ^{-s}y^{t}\mathrm{d}y\quad\!\!\!\!\mathrm{d}s\quad\!\!\!\!\mathrm{d}t,
\end{eqnarray}
where  $i^2=-1$, and $\mathcal{C}_1$ and $\mathcal{C}_2$ denote the $s$ and $t$-planes, respectively. Finally, simplifying $\int_{0}^{\infty}\left( 1+\frac{\cal C}{y}\right) ^{-s}y^{t}\mathrm{d}y$ to $\frac{{\cal C}^{1+t}\Gamma(-1-t)\Gamma(1+t+s)}{\Gamma(s)}$ by means of \cite[Eqs. (8.380.3) and (8.384.1)]{grad}  while utilizing the relations $\Gamma(1-rs)=-rs\Gamma(-rs)$, and $s\Gamma(s)=\Gamma(1+s)$ then \cite[Eq. (1.1)]{mittal} yield  (\ref{eqC2:CDFfix}).
\section{Appendix B\\Proof of Lemma 1}
\label{app:B}
Resorting to the Mellin-Barnes representation of the bivariate Fox-H function \cite[Eq. (2.57)]{mathai} in (\ref{eqC2:CDFfix}) yields
 \begin{eqnarray}
 \label{eqapp:1}
P_{\text{out}}\!\!&=&\!\frac{\xi^2A\kappa m {\cal C} }{\Gamma(\alpha)\Gamma(Nm)\Gamma(\kappa)\Gamma(Lm_I)\Gamma(\kappa_I)\kappa_I m_I \bar{\gamma}}\!\!\! \nonumber\\
 \!\!\!\!\!\!\!\!\!\!&&\!\!\!\!\!\!\!\!\!\!\!\!\!\!\!\!\!\!\!\!\!\!\! \sum_{k=1}^{\beta}\!\! \frac{b_k}{\Gamma(k)}\frac{1}{4\pi^2i^2}\int_{\mathcal{C}_1}^{}\!\!\int_{\mathcal{C}_2}^{}\!\!\!\!\frac{\Gamma(\xi^2+rs)\Gamma(k+rs)\Gamma(\alpha+rs)}{\Gamma(\xi^2+1+rs)\Gamma(1+s)}\nonumber\\
 \!\!\!\!\!\!\!\!\!\!&&\!\!\!\!\!\!\!\!\!\!\!\!\!\!\!\!\!\!\!\!\!\!\!\times \frac{\Gamma(-1-t)^2}{\Gamma(1+t)} \frac{\Gamma(\kappa-1-t)\Gamma(N m-1-t)\Gamma(2+t)}{\Gamma(-t)}\!\!\!\nonumber\\
 \!\!\!\!\!\!\!\!\!\!&&\!\!\!\!\!\!\!\!\!\!\!\!\!\!\!\!\!\!\!\!\!\!\!\times\Gamma(1+\kappa_I+t)\Gamma(1+Lm_I+t) \Gamma(1+s+t)\nonumber\\
 \!\!\!\!\!\!\!\!\!\!&&\!\!\!\!\!\!\!\!\!\!\!\!\!\!\!\!\!\!\!\!\!\!\! \left(\!\frac{\kappa m{\cal C}}{\kappa_Im_I\bar{\gamma}}\!\right)^{\!\!t}\left(\!\frac{B^r \gamma_{th}}{\mu_r}\!\right)^{\!\!-s}\mathrm{d}s\quad\!\!\!\!\mathrm{d}t,\nonumber\\
 \!\!\!\!\!\!\!\!\!\!\!\!\!\!\!\!\!\!\!\!&\!\!\!\!\!\!\!\!\!\!\!\!\!\!\! \!\!\!\!\!\!\!\!\!\!\overset{(a)}{=}& \!\!\!\!\!\!\!\!\!\!\!\!\!\!\!\!\frac{\xi^2A\kappa m {\cal C} }{\Gamma(\alpha)\Gamma(Nm)\Gamma(\kappa)\Gamma(Lm_I)\Gamma(\kappa_I)\kappa_I m_I \bar{\gamma}}\sum_{k=1}^{\beta}\!\! \frac{b_k}{\Gamma(k)} \nonumber\\
 \!\!\!\!\!\!\!\!\!\!&&\!\!\!\!\!\!\!\!\!\!\!\!\!\!\!\!\!\!\!\!\!\!\! \frac{1}{2\pi i}\int_{\mathcal{C}_2}^{} \frac{\Gamma(-1-t)^2}{\Gamma(1+t)} \frac{\Gamma(\kappa-1-t)\Gamma(N m-1-t)}{\Gamma(-t)}\!\!\!\nonumber\\
 \!\!\!\!\!\!\!\!\!\!&&\!\!\!\!\!\!\!\!\!\!\!\!\!\!\!\!\!\!\!\!\!\!\!\times\Gamma(2\!+\!t)\Gamma(1\!+\!\kappa_I\!+\!t)\Gamma(1\!+\!Lm_I+t) \left(\!\frac{\kappa m{\cal C}}{\kappa_Im_I\bar{\gamma}}\!\right)^{\!\!t}\nonumber\\
 \!\!\!\!\!\!\!\!\!\!&&\!\!\!\!\!\!\!\!\!\!\!\!\!\!\!\!\!\!\!\!\!\!\!\times{\rm H}_{2,4}^{4,0}\Biggl[\!\!{\frac{B^r \gamma_{th}}{\mu_r}} \!\left|\begin{array}{cccc} \!\!(1,1),(1+\xi^2,r)\\ \!\! (1\!+\!t,1),(\alpha,r),(k,r),(\xi^2,1)\end{array}\right.\!\!\!\!\!\Biggr]\mathrm{d}t,
 \end{eqnarray}
where (a) follows from using the definition of the $\rm H$-function shown in \cite[Eq. (1.1.1)]{kilbas}. Therefore, by applying \cite[Theorem 1.11]{kilbas} to  (\ref{eqapp:1}) when $\mu_r/\gamma_{th}\rightarrow\infty$ yields after some algebraic manipulations
\begin{eqnarray}
		P_{\text{out}}&\underset{\frac{\mu_r}{\gamma_{th}}\gg1}{\approx}&\frac{\xi^2A \frac{\kappa m}{\kappa_Im_I} {\cal C}}{\Gamma(\alpha)\Gamma(Nm)\Gamma(\kappa)\Gamma(Lm_I)\Gamma(\kappa_I) \bar{\gamma}_2}\sum_{k=1}^{\beta} \frac{b_k}{\Gamma(k)} \nonumber\\
		&&\Biggl(\frac{\Gamma(\alpha-\xi^2)\Gamma(k-\xi^2)}{r \Gamma(1-\frac{\xi^2}{r})}\quad\Xi'\left(\gamma_{th},\frac{\xi^2}{r}\right) \nonumber\\
		&&+\frac{\Gamma(\xi^2-\alpha)\Gamma(k-\alpha)}{r \Gamma(1-\frac{\alpha}{r})\Gamma(1+\xi^2-\alpha)}\quad\Xi'\left(\gamma_{th},\frac{\alpha}{r}\right) \nonumber\\
		&&+\frac{\Gamma(\xi^2-k)\Gamma(\alpha-k)}{r \Gamma(1-\frac{k}{r})\Gamma(1+\xi^2-k)}\quad\Xi'\left(\gamma_{th},\frac{k}{r}\right) \nonumber\\
		&&+\frac{B^r\gamma_{th}}{\mu_r}{\rm H}_{3,5}^{5,2}\Biggl[{\frac{\kappa m{\cal C}B^r\gamma_{th}}{\kappa_Im_I\bar{\gamma}_2\mu_r}}\left|\begin{array}{cccc}(\sigma,\Sigma)\\(\phi,\Phi)\end{array}\right.\!\!\! \Biggr]\!\Biggr),
	\end{eqnarray}
where $(\sigma,\Sigma)=(-\kappa_I,1),(-Lm_I,1),(1+\xi^2-r,r)$, $(\phi,\Phi)=(\xi^2-r,r),(\alpha-r,r),(k-r,r),(\kappa-1,1),(Nm-1,1)$, and
 $\Xi'(x,y)=\left(\frac{B^rx}{\mu_r}\right)^{y}\!\!{\rm G}_{5,5}^{4,4}\!\!\left[\!\!{\frac{\kappa m{\cal C}}{\kappa_I m_I\bar{\gamma}_2}}\!\!\left|  \begin{array}{cccc}-\kappa_I\!,-\!Lm_I\!,\!-1,y,0\!\\ \kappa\!-\!1,Nm-1,\!-1,\!-1,0\end{array}\!\!\right.\right]$.  Finally applying \cite[Eq. (931.5)]{grad} completes the proof.

\section{Appendix C\\Proof of Lemma 4}
\label{app: C}
From \cite{imen2}, the average capacity can be computed as
\begin{equation}
\label{eq1:7}
C=\frac{1}{2\ln (2)} \displaystyle\int_0^\infty se^{-s}M^{(c)}_{\gamma_1}(s)M^{(c)}_{\gamma_2}(s)ds,
\end{equation}	
where $M^{(c)}_X(s)=\int_{0}^{\infty}e^{-sx}F^{(c)}_X(x) dx$ stands for the complementary MGF (CMGF).
The CMGF of the first hop's SNR $\gamma_1$ under M\'alaga-$\mathcal{M}$ distribution with pointing errors  is given by \cite[Eq. (9)]{Trigui}
\begin{equation}
\label{eqC2:apD1}
M^{(c)}_{\gamma_1}(s)=\frac{\xi^2A  r  \mu_r}{\Gamma(\alpha)B^r}\sum_{k=1}^{\beta} \frac{b_k}{\Gamma(k)}{\rm \mathcal{H}}_{4,3}^{1,4} \Biggl[\!\frac{\mu_r}{B^r} s\Bigg\vert \  {(\delta_2,\Delta_2)\atop (\lambda_2,\Lambda_2)}\Biggr].
\end{equation}
Moreover, the Laplace transform of the RF link's CCDF yields its CMGF after resorting to \cite[Eq. (7.813.1)]{grad} and \cite[Eq. (1.111)]{mathai} as
\begin{equation}
\label{eqC2:apD2}
M^{(c)}_{\gamma_2}(x)=\frac{{\rm \mathcal{H}}_{3,4}^{3,3}\Biggl[\frac{\kappa_Im_I\bar{\gamma}}{\kappa m} s\Bigg\vert\  {(\chi_2,X_2)\atop(\upsilon_2,\Upsilon_2)}\Biggr] }{s\Gamma(Nm)\Gamma(\kappa)\Gamma(Lm_I)\Gamma(\kappa_I)}.
\end{equation}
Finally, (\ref{eq:18}) follows after plugging (\ref{eqC2:apD1}) and (\ref{eqC2:apD2}) into (\ref{eq1:7}) and applying \cite[Eq. (2.2)]{mittal}.

\end{document}